\documentclass[prb,
12pt,
superscriptaddress,showpacs,amsmath,amssymb]{revtex4}
\usepackage{amsfonts}
\usepackage{bm}
\usepackage{verbatim}

\usepackage{graphicx}

\begin{document}

\author{G.E.~Volovik}
\affiliation{Low Temperature Laboratory, Aalto University,  P.O. Box 15100, FI-00076 Aalto, Finland}
\affiliation{Landau Institute for Theoretical Physics, acad. Semyonov av., 1a, 142432,
Chernogolovka, Russia}

\title{Dimensionless physics:  Planck constant as an element of Minkowski metric}

\date{\today}

\begin{abstract}
Diakonov theory of quantum gravity, in which tetrads emerge as the bilinear combinations  of the fermionis fields,\cite{Diakonov2011} suggests  that in general relativity  the metric may have dimension 2, i.e. $[g_{\mu\nu}]=1/[L]^2$.  Several other approaches to quantum gravity, including the model of superplastic vacuum  and $BF$-theories of gravity support  this suggesuion. The important consequence of such metric dimension is that all the diffeomorphism invariant quantities are dimensionless for any dimension of spacetime. These include the action $S$, interval $s$, cosmological constant $\Lambda$, scalar curvature $R$, scalar field $\Phi$, etc.  
Here we are trying to further exploit the Diakonov idea, and consider the dimension of the Planck constant. The application of the Diakonov theory suggests that the Planck constant $\hbar$ is the parameter of the Minkowski metric.
The Minkowski parameter $\hbar$ is invariant only under Lorentz transformations, and is not diffeomorphism invariant.  As a result the Planck constant $\hbar$ has nonzero dimension -- the dimension of length [L].
Whether this Planck constant length is related to the Planck length scale, is an open question.
 In principle there can be different Minkowski vacua with their own values of the parameter $\hbar$. Then in the thermal contact between the two vacua  their temperatures obey the analog of the Tolman law:  $\hbar_1/T_1= \hbar_2/T_2$. 
\end{abstract}

\maketitle

\newpage

\section{Introduction. }
\subsection{Dimensional tetrads}

Several approaches to quantum gravity (Diakonov tetrads emerging  as the bilinear combinations  of the fermionis fields;\cite{Diakonov2011} the model of superplastic vacuum; $BF$-theories of gravity; and effective acoustic metric)  suggest\cite{Volovik2021,Volovik2021cont}  that in general relativity  the metric must have dimension 2, i.e. $[g_{\mu\nu}]=1/[L]^2$.
In particular, the model of the superplastic vacuum\cite{KlinkhamerVolovik2019} is
 described in terms of the so-called elasticity tetrads:\cite{Dzyaloshinskii1980,NissinenVolovik2019,Nissinen2020,Nissinen2020a,Nissinen2020b,Burkov2021}
\begin{equation}
E^a_\mu= \frac{\partial X^a}{\partial x^\mu}\,,
\label{ElasticityTetrads}
\end{equation}
where equations $X^a(x)=2\pi n_a$ are equations of the (deformed) crystal planes. Since  the functions $X^a$ play the role of the geometric $U(1)$ phases and thus are dimensionless, the elasticity tetrads play the role of the gauge fields (translation gauge fields). That is why tetrads have the same dimension 1 as the dimension of gauge fields:
\begin{equation}
[E^a_\mu] = \frac{1}{[L]}\,.
\label{ElasticityTetradsDim}
\end{equation}
The dimension $n$ of quantity $A$ means  $[A]=[L]^{-n}$, where $[L]$ is dimension of length.   The $U(1)$ gauge fields, which form the dimensional tetrads, are not necessarily related to translations and the crystalline structure of the vacuum. The latter is only one of the possible sources of the dimensional tetrads. The dimension 1 tetrads appeared first in the Diakonov theory,\cite{Diakonov2011,VladimirovDiakonov2012,VladimirovDiakonov2014,ObukhovHehl2012} where the tetrads emerge as bilinear combinations  of the quantum fermionic fields:
\begin{equation}
\hat E^a_\mu = \frac{1}{2}\left( \Psi^\dagger \gamma^a\partial_\mu  \Psi -  \Psi^\dagger\overleftarrow{\partial_\mu}  \gamma^a\Psi\right) \,.
\label{TetradsFermions}
\end{equation}
This theory suggests that the metric and thus the space-time distance are both the quantum objects made of the  leptons and quarks.\cite{Akama1978} 

In Diakonov theory the fermionic fields $\Psi$ are dimensionless,\cite{VladimirovDiakonov2012} and thus the dimension of the fermionic tetrads is the same as the dimension of the elasticity tetrads in Eq.(\ref{ElasticityTetradsDim}). The fermionic tetrads in Eq.(\ref{TetradsFermions}) serve as the quantum spinor generalization of the elasticity tetrads. In Eq.(\ref{TetradsFermions}) tetrads are operators, but in what follows we consider tetrads as the vacuum expectation values of these operators, $E^a_\mu=<\hat E^a_\mu>$.

\subsection{Dimensional metric} 

Fermionic tetrads in Eq.(\ref{TetradsFermions})  and elasticity tetrads in Eq.(\ref{ElasticityTetrads}) andgive rise to the metric with lower indices, which is the following bilinear combination of tetrads:
\begin{equation}
g_{\mu\nu}=\eta_{ab}E^a_\mu E^b_\nu \,.
\label{MetricElasticity}
\end{equation}
This covariant metric tensor has dimension $n=2$, while the contravariant metric $g^{\mu\nu}$ has dimension $n=-2$:
 \begin{equation}
[g_{\mu\nu}] =\frac{1}{[L]^2}\,\,,\,\, [g^{\mu\nu}] =[L]^2\,.
\label{MetricDimension}
\end{equation}

The determinant of the tetrads has dimension 4 in the 4-dimensional spacetime:
 \begin{equation}
[e]=[\sqrt{-g}\,] =\frac{1}{[L]^4} \,.
\label{DimensionDeterminant}
\end{equation}
In the $N$-dimensional spacetime, the dimensions of the metric elements are the same as in Eq.(\ref{MetricDimension}), while the tetrad  determinant has dimension $N$:
 \begin{equation}
[e]=[\sqrt{-g}\,] =\frac{1}{[L]^N} \,.
\label{DimensionDeterminantN}
\end{equation}

Eq.(\ref{MetricElasticity}) gives rise to the dimensionless interval:
 \begin{equation}
ds^2=g_{\mu\nu}dx^\mu dx^\nu \,\,, \,\,  [s^2]= \frac{1}{[L]^2} \cdot [L]^2=[1] \,.
\label{DimensionInterval}
\end{equation}

 What is the consequences of such unusual dimension of metric? Does such metric describe geometry or dynamics?
 Let us consider this on example of dynamics of classical point particle.

\section{Dynamics of classical point particle and dimensionless physics}

The interval in Eq.(\ref{DimensionInterval}) describes in particular the classical dynamics of a point particle with action:
 \begin{equation}
S=M\int ds \,.
\label{particleAction}
\end{equation}
Here the particle mass  $M$ in  Eq.(\ref{particleAction}) is dimensionless, $[M]=[1]$, as well as all other diffeomorphism invariant quantities, such as action $S$, interval $s$, cosmological constant $\Lambda$, scalar curvature $R$, scalar field $\Phi$, etc.\cite{Volovik2021,Volovik2021cont} This is valid for the arbitrary dimension $N$ of spacetime, and thus is universal,
which is the most important consequence of the metric dimension $1/[L]^2$.

The variation of action leads to the Hamilton--Jacobi equation expressed in terms of the contravariant metric $g^{\mu\nu}$:
 \begin{equation}
g^{\mu\nu}\partial_\mu S \partial_\nu S + M^2=0 \,.
\label{HJ}
\end{equation}
Both terms in Eq.(\ref{HJ}) are dimensionless. 

Since mass is dimensionless, such quantities as $e^2/r$ and $GM^2/r$ are dimensionless, which leads to:
 \begin{equation}
[e^2]=[G]=[L] \,.
\label{e2G}
\end{equation}

\section{Minkowski spacetime and Planck constant}

Let us introduce the wave vector:
 \begin{equation}
 k_\mu=\partial_\mu S \,\, , \,\, [k_\mu]=\frac{1}{[L]}\,.
\label{WV}
\end{equation}
It has dimension of inverse length and obeys equation:
 \begin{equation}
g^{\mu\nu}k_\mu k_\nu  + M^2=0 \,.
\label{WaveVector}
\end{equation}
Let us now consider this equation  in the flat Minkowski spacetime, where the interval becomes:
\begin{equation}
 ds^2_{\rm Mink}= |g_{00}^{\rm Mink}| (-dt^2 + d{\bf r}^2) \equiv \frac{1}{{\hat h}^2} (-dt^2 + d{\bf r}^2)\,.
\label{conformal_hbar}
\end{equation}
Here we introduced the parameter ${\tilde h}\equiv 1/\sqrt{-g_{00}^{\rm Mink}}$ with dimension of length,  $[{\tilde h}]=[L]$.
In Minkowski spacetime the wave vector obeys equation:
 \begin{equation}
{\tilde h}^2(\omega^2 -{\bf k}^2) = M^2\,.
\label{WaveVector2}
\end{equation}
This suggests that it is natural to identify $\tilde h$ with Planck constant:\cite{Volovik2009}
 \begin{equation}
\hbar\equiv{\tilde h}=\frac{1}{\sqrt{-g_{00}^{\rm Mink}}} \,.
\label{hbar}
\end{equation}

In general, the dimensional metric leads to the difference between  the dimensional frequency, $[\omega_{mn}]=1/[L]$, and the  dimensionless mass and energy:
\begin{equation}
 [E]=[M]=[\sqrt{g^{00}}] [\omega] =[L] \cdot  \frac{1}{ [L]} =[1] \,.
\label{FrequencyDimension}
\end{equation}
The red shift equation, $M_m - M_n=\sqrt{-g^{00}}\, \omega_{mn}$, is the generalization of the conventional relation between the energy levels and frequency of radiating photon in Minkowski vacuum, $E_m -E_n=\hbar \omega_{mn}$. This supports the suggestion that $\hbar$ is the constant conformal factor  in a given Minkowski vacuum, $\hbar\equiv \sqrt{-g^{00}_{\rm Mink}}$, with dimension of length, $[\hbar]=[L]$. As a results the energy in equation $E=\hbar \omega$ is dimensionless,
$[E] = [\hbar] [\omega] =[1]$.

Note that as distinct from the dimensionless  quantities, which  are  diffeomorphism invariant, the parameter $\hbar$ is not diffeomorphism invariant. It is determined only in the Minkowski vacuum, and being the element of the Minkowski metric it is invariant only under Lorentz transformations. As a result the Planck constant $\hbar$ is not dimensionless, and has the dimension of length [L].
Then according to Weinberg criterion \cite{WeinbergCriterium}  $\hbar$ cannot be the fundamental constant  (see also Refs. \cite{Trialogue,OkunCube,OkunGamov,Gamov} on fundamental constants).

Another parameter of Minkowski spacetime is the speed of light $c$, which enters the metric in the following way: $g^{\mu\nu}_{\rm Mink}={\rm diag} (\hbar^2, \hbar^2 c^2,\hbar^2 c^2, \hbar^2 c^2)$. This parameter is invariant only under space rotation group $SO(3)$, and has dimension $[L]/[t]$. If the parameter $c$ is taken into account, the Planck constant has dimension of time,  $[\hbar] = [M][t]=[t]$.
Then according to the dimension of electric charge in Eq.(\ref{e2G}), the fine structure constant $e^2/\hbar c$ remains dimensionless. 
In what follows, except for the special cases, we do not consider the parameter $c$ and use units with $c=1$.

The parameter $\hbar$ enters only the Minkowski metric, and does not enter any equation written in the covariant form, i.e. in terms of the full metric. That is why in general the commutation relations for position and momentum operators in quantum mechanics do not contain $\hbar$:
\begin{equation}
 [{\hat k}_i ,{\hat x}^j]=i \delta_i^j \,.
\label{Commutation}
\end{equation}
As a result, the elementary volume of phase space $\int dk\, dx$ and also the action are dimensionless,
while the parameter $\hbar$ of Minkowski vacuum has dimension of time $[t]$.

One can also introduce the momentum $p^\mu$:
 \begin{equation}
 p^\mu=g^{\mu\nu}k_\nu  \,\, , \,\,   g_{\mu\nu} p^\mu p^\nu+ M^2=0\,.
\label{momentum}
\end{equation}
In  Minkowski  spacetime one obtains:
 \begin{equation}
 p^\mu=\hbar^2 k_\nu \,\, ,\,\, [p^\mu]=[L]\,.
\label{momentumMinkowski}
\end{equation}
As distinct from the momentum in the vacua described by the conventional dimensionless metric, this momentum contains $\hbar^2$ and has dimension $[L]$ instead of $1/[L]$.

\section{Schr\"odinger equation in Minkowski spacetime}

Let us consider the quadratic terms in the action for the classical scalar field $\Phi$ in the $N$-dimensional spacetime:
\begin{equation}
S=\int d^N x\,\sqrt{-g} \,\left(  g^{\mu\nu} \nabla_\mu \Phi^*  \nabla_\nu \Phi +M^2|\Phi|^2 \right)\,,
\label{scalar}
\end{equation}
Taking into account equations (\ref{MetricDimension}) and (\ref{DimensionDeterminantN}), one obtains that the scalar field is dimensionless, $[\Phi]=[1]$, for arbitrary spacetime dimension $N$. This universal zero dimension differs from the $N$-dependent dimension of scalar fields in the conventional approach, where the dimension is   $n=(N-2)/2$.

 Expanding the Klein-Gordon equation for scalar $\Phi$ in Eq.(\ref{scalar}) over $1/M$ one obtains the non-relativistic Schr\"odinger action.  In Minkowski spacetime, introducing the Schr\"odinger wave function $\psi$:
\begin{equation}
\Phi({\bf r},t) = \frac{1}{\sqrt{M}}\exp\left(i Mt /\sqrt{-g^{00}}\right)\psi({\bf r},t)  \,,
\label{eq:PhiPsi}
\end{equation}
 one obtains the Schr\"odinger-type  action in the form
\begin{eqnarray}
S_\text{Schr}=\int d^3x dt  \sqrt{-g}\, {\cal L} \,,
\label{eq:SchroedingerAction}
\\
2{\cal L}= 
i\sqrt{-g^{00}} \left(\psi \partial_t \psi^*-\psi^* \partial_t \psi\right) 
+\frac{g^{ik}}{M}\nabla_i\psi^* \nabla_k \psi   \,.
\label{eq:SchroedingerEq}
\end{eqnarray}

Since in Minkowski vacuum the metric elements play the role of $\hbar$:
 \begin{equation}
\sqrt{-g^{00}_{\rm Mink}} \equiv\hbar\,\,,\,\, g^{ik}_{\rm Mink}\equiv\hbar^2 \delta^{ik}\,,
\label{MinkowskiMetric}
\end{equation}
one obtains the conventional Schr\"odinger  Lagrangian and the Schr\"odinger wave equation:
\begin{equation}
i\hbar \partial_t \psi =-\frac{\hbar^2}{2M} \nabla^2\psi \,.
\label{SchrodingerEq}
\end{equation}
This is another consequence of the metric with dimension $1/[L]^2$: the quantum mechanical Schr\"odinger equation  for nonrelativistic particle  is obtained directly from the classical relativistic scalar field. 

The same equation  for nonrelativistic fermions can be obtained from classical action for massive Dirac particles:
\begin{equation}
S=\int d^4x\,  e\, (ie^\mu_a \bar\Psi  \gamma^a \nabla_\mu \Psi - M\bar\Psi \Psi)\,,
\label{Fermions}
\end{equation}
where $e^\mu_a$ are tetrads that are inverse to $E_\mu^a$, with dimension of length, $[e^\mu_a]=[L]$.
In the Minkowski  spacetime one has
\begin{equation}
 e^\mu_a =\hbar\, {\rm diag}(-1,1,1,1)\,,
\label{MinkowskiTetrad}
\end{equation}
and in the limit $\hbar |{\bf k}|\ll M$ one obtains Eq.(\ref{SchrodingerEq}).

The corresponding Hamiltonian for nonrelativistic particle is dimensionless together with energy $E$:
\begin{equation}
{\cal H}= -\frac{\hbar^2}{2M}\nabla^2 \,\,, \,\, [{\cal H}]  =[\hbar]^2 \cdot [\nabla]^2\cdot [M]^{-1}=[L]^2 \cdot  \frac{1}{ [L]^2} \cdot [1] =[1] \,.
\label{HamiltonianEq}
\end{equation}

If to follow the elasticity tetrad scenario in Eq. (\ref{ElasticityTetrads}), then $\hbar$ with its dimension of length, 
$[\hbar]=[L]$, plays the role of the "interatomic" distance in the medium called the relativistic quantum vacuum. 
Then  the Schr\"odinger equation (\ref{SchrodingerEq}) can be valid only in the long wavelength limit, $\hbar |{\bf k}| \ll M \ll 1$. 
\section{de Sitter spacetime and Planck constant}

The same conformal factor in terms of $\hbar$, which enters Eq.(\ref{conformal_hbar}), is valid for any $D+1$ Minkowski spacetime including the 4+1 Minkowski spacetime, which determines the de Sitter (dS) spacetime:
\begin{equation}
 g_{\rm dS}^{\mu\nu}X^\mu X^\nu=\alpha^2\,.
\label{dS}
\end{equation}
Here $\alpha$ is dimensionless constant, and $g_{\rm dS}^{\mu\nu}=1/\hbar^2$. This equation gives the Hubble parameter:
\begin{equation}
H=\frac{1}{\hbar \alpha}\,\,, \,\, [H]=\frac{1}{[L]}\,.
\label{dS_H}
\end{equation}
The dS metric in the Paineve-Gullstrand form contains two parameters, $\hbar$ and $H$ (actually three parameters if the speed of light is included):
\begin{equation}
ds^2 =\frac{1}{\hbar^2}\left( -dt^2 + (dr -Hrdt)^2 + r^2 d\Omega^2 \right)\,.
\label{dSds}
\end{equation}
At $r=0$ the metric is Minkowski, however it is not excluded that in the dS universe the parameter $\hbar$ deviates from its value in Minkowski vacuum and depends on $\alpha$.

The probability of Hawking radiation of particle with mass $M$ detected by observer at $r=0$ is
\begin{equation}
w \propto \exp\left(- \frac{2\pi  M}{H \sqrt{-g^{00}_{\rm Mink}}} \right)=\exp\left(- \frac{2\pi  M}{\hbar H} \right)\,.
\label{HawkingdS}
\end{equation}

\section{Planck constant and Tolman law}

Eq.(\ref{HawkingdS}) corresponds to the following Hawking temperature:
\begin{equation}
T(r=0)=\frac{ \sqrt{-g^{00}_{\rm Mink}} H}{2\pi}=\frac{\hbar H}{2\pi} \,.
\label{HawkingTdS}
\end{equation}
The Hawking temperature is dimensionless, $[T(r=0)] = [L] \cdot 1/[L]=[1]$.
Note that constant $H/2\pi$ plays the role of Tolman temperature, which enters the Tolman law:
\begin{equation}
T(r)=\frac{T_{\rm Tolman}}{ \sqrt{-g_{00}(r)} } \,\,, \,\,  T_{\rm Tolman}=\frac{H}{2\pi}\,.
\label{HawkingTolman}
\end{equation}

While temperature is dimensionless, $[T]=[M]$, the Tolman temperature has dimension of inverse length (or of inverse time, if the speed of light $c$ is not ignored): $[T_{\rm Tolman}]=[H]=1/[L]$, see also Refs.\cite{Volovik2021,Volovik2021cont}.

The parameter $\hbar$ determines the ratio between the temperature and Tolman temperature in the Minkowski vacuum. In principle there can be different Minkowski vacua with their own values of $\hbar$.\cite{Klinkhamer2022b} Then the Tolman law  can be applied to the boundary between the two Minkowski vacua with different values of $\hbar$. In thermal equilibrium they must have the same Tolman temperature, and thus their temperatures obey the rule, $\hbar_1/T_1= \hbar_2/T_2$. This means that in thermal equilibrium the contacting Minkowski vacua  have the same time on imaginary axis, $\tau_1=\tau_2$.

Snce the Minkowski metric is quadratic in $\hbar$, the parameter $\hbar$ may have negative sign, which corresponds to the different signs of the tetrad elements in Minkowski vacuum. That is why, in the above equations the $\hbar$ should be substituted by the modulus $|\hbar|$. In principle, there can be the cosmological domain walls between the Minkowski vacua with positive and negative sign of $\hbar$ or/and $c$.\cite{Volovik2009} Example of such wall can be found in Ref.\cite{Vergeles2022}.

\section{Length dimension of Newton constant and Planck length}

In previous works we considered the Newton constant $G$ as dimensionless.\cite{Volovik2021,Volovik2021cont} This is because we  did not take into account the dimension of $\hbar$. Now, since the Planck constant has dimension of length, the Newton constant also acquires the dimension of length, $[G]=[L]$ in Eq.(\ref{e2G}). The gravitational action is  dimensionless and has the following form:
\begin{eqnarray}
\hspace*{-10mm}
S = \int d^4x\,\sqrt{-g} \, \left( \Lambda +\frac{1}{16\pi}M_{\rm P}^2\, R  + \beta {\cal R}^2 + ...\right) + \gamma \int \,E^a \wedge E^b\wedge R_{ab} \,.
\label{EinsteinAction4D}
\end{eqnarray}
Here the $\beta  {\cal R}^2$ means all the terms which are quadratic in curvature tensor ${\cal R}_{\alpha\beta\mu\nu}$, and the last term is Barbero-Immirzi action, where $E^a=E^a_\mu dx^\mu$ is the 1-form tetrad, and $R_{ab}$ is the curvature 2-form. The cosmological constant $\Lambda$ and  scalar curvature $R$ are dimensionless, 
$[\Lambda]=[R]=[1]$, as well as all ${\cal R}^2$ terms where the parameters $\beta$ are dimensionless. The parameter $\gamma$, which enters the Barbero-Immirzi  term, is also dimensionless due to Eq.({\ref{ElasticityTetradsDim}). The parameter $M_{\rm P}=\sqrt{\hbar/G}$ is the Planck mass, which is dimensionless as all the masses $M$ in this approach. Here the length dimension of the Newton constant is compensated by the length dimension the Planck constant, $[M_{\rm P}]^2=[\hbar]/[G]=[L] [L]^{-1}=[1]$. 

 Both the Planck constant and the Newton constant do not enter Eq.(\ref{EinsteinAction4D}), which supports the view, that neither of them is the fundamental constant. On the other hand it is not excluded that the dimensionless quantity $M_{\rm P}^2$ can be the fundamental constant, which is determined by symmetry and topology. The same concerns dimensionless parameter $\gamma$ in the  Barbero-Immirzi term, as well as the prefactor in the topological Nieh-Yan term,\cite{NiehYan1982a,NiehYan1982b,Nieh2007} which is also dimensionless in this approach.\cite{NissinenVolovik2019} 
  
 The Planck length scale has the conventional form 
$l^2_{\rm P}= \hbar G$, with $[l_{\rm P}]^2= [\hbar] [G]=[L] [L]=[L]^2$.
The Planck constant has the same dimension as the Planck length, $[\hbar]=[l_{\rm P}]=[L]$. Whether this "Planck constant length" is related to the "Planck length scale", is an open question.\cite{Carlip2022} Anyway, this suggests the close connection between gravity and quantum mechanics. 

The dimensionless cosmological constant $\Lambda$ can be expressed in terms of the vacuum energy density,
$\Lambda= \hbar^4 \epsilon_{\rm vac}= (g_{00}^{\rm Mink})^{-2}\epsilon_{\rm vac}$. This looks similar to the Adler proposal of the Weyl scaling invariant form of the dark energy action,\cite{Adler2022,Adler2022b}  where the vacuum energy density 
$(g_{00})^{-2}\epsilon_{\rm vac}$ contains the metric element $g_{00}$ instead of its Minkowski value $g_{00}^{\rm Mink}$.
However, if the parameter $c$ in  Minkowski metric  $g^{\mu\nu}_{\rm Mink}={\rm diag} (\hbar^2, \hbar^2 c^2,\hbar^2 c^2, \hbar^2 c^2)$ is not ignored, the vacuum energy can be written in the following form:
\begin{eqnarray}
S_\Lambda = \int d^4x\,\frac{\sqrt{-g}}{\sqrt{-g^{\rm Mink}}}\, \epsilon_{\rm vac} \,.
\label{VacuumAction}
\end{eqnarray}
This looks similar to the bi-metric approach  with the so-called prior metric.\cite{Klinkhamer2017}
Note that due to its dimension, $\sqrt{-g^{\rm Mink}}$ may play the role of the equilibrium 4D density of “spacetime atoms” \cite{Klinkhamer2022} or the equilibrium density of the lattice points in the self-sustained quantum vacuum.\cite{Klinkhamer2022b}

\section{Conclusion}

The important consequence of the Diakonov theory, in which the metric dimension is $1/[L]^2$, is that all the diffeomorphism invariant quantities are dimensionless for any dimension of spacetime. These include the action $S$, interval $s$, cosmological constant $\Lambda$, scalar curvature $R$, scalar field $\Phi$, Planck mass $M_{\rm P}$, masses of particles and fields, etc.  

Another consequence of the metric dimension $1/[L]^2$, is that the Planck constant $\hbar$ is the property of the Minkowski vacuum.  Being the element of the Minkowski metric, $\hbar$ is invariant only under Lorentz transformations, and is not diffeomorphism invariant.  As a result the Planck constant $\hbar$ is not dimensionless, and has the dimension of length $[L]$, the same dimension as the Newton constant $G$.
Whether this Planck constant length $\hbar$ is related to the Planck length scale $ \sqrt{\hbar G}$ is an open question.
It is possible that the dimensionless quantity $\hbar/G$ can be the  fundamental constant, which reflects the symmetry or/and topology of the quantum vacuum.\cite{Volovik2021cont} 

 In principle it is not excluded that there can be different Minkowski vacua, with cosmological phase transitions between these vacua.\cite{Klinkhamer2022b} Then each  vacuum may have its own value of the parameter $\hbar$. In this case the thermal contact between the two vacua obeys the Tolman law:  in thermal equilibrium their temperatures are connected in the following way, $\hbar_1/T_1= \hbar_2/T_2$.

All these consequences of the consequences of Diakonov theory suggest that metric describes the dynamics, quantum mechanics and thermodynamics, rather than the geometry.

 {\bf Acknowledgements}.  I thank Y.N. Obukhov and N.N. Nikolaev for discussion and criticism. This work has been supported by the European Research Council (ERC) under the European Union's Horizon 2020 research and innovation programme (Grant Agreement No. 694248).


\begin{thebibliography}{99}

\bibitem{Diakonov2011}
D. Diakonov,
arXiv:1109.0091.

\bibitem{Volovik2021} 
G.E. Volovik,
ZhETF {\bf 159}, 815--821 (2021),
JETP {\bf 132},  727--733 (2021).

\bibitem{Volovik2021cont} 
G.E. Volovik,
ZhETF {\bf 162},  680--685 (2022).

\bibitem{KlinkhamerVolovik2019}
F.R. Klinkhamer and G.E. Volovik,
Pis'ma ZhETF  {\bf 109}, 369--370 (2019),
JETP Lett. {\bf 109},  364--367 (2019).

\bibitem{Dzyaloshinskii1980}
I.E. Dzyaloshinskii and G.E. Volovick, 
Ann. Phys. {\bf 125}, 67--97 (1980).

\bibitem{NissinenVolovik2019} 
J. Nissinen and G.E. Volovik,
Phys. Rev. Research {\bf 1}, 023007 (2019).

\bibitem{Nissinen2020} 
J. Nissinen, 
Ann. Phys. {\bf 447}, 169139 (2022).

\bibitem{Nissinen2020a} 
J. Nissinen, 
Phys. Rev. Lett. {\bf 124}, 117002 (2020).

\bibitem{Nissinen2020b} 
S. Laurila and J. Nissinen,
Phys. Rev. B {\bf 102}, 235163 (2020).
 
\bibitem{Burkov2021} 
L. Gioia, Chong Wang, and A.A. Burkov,
Phys. Rev. Research {\bf 3}, 043067 (2021).


\bibitem{VladimirovDiakonov2012}
A.A. Vladimirov and D. Diakonov,
Phys. Rev. D {\bf 86}, 104019 (2012).

 \bibitem{VladimirovDiakonov2014}
A.A. Vladimirov and D. Diakonov,
Physics of Particles and Nuclei {\bf 45}, 800 (2014).
 
\bibitem{ObukhovHehl2012}
Y.N. Obukhov and F.W. Hehl,
Phys. Lett. B {\bf 713}, 321--325 (2012).

\bibitem{Akama1978}
 K. Akama, 
PTP {\bf 60}, 1900--1909 (1978).
 
\bibitem{Volovik2009} 
G.E. Volovik,  
JETP Lett. {\bf 90}, 697--704 (2009).

\bibitem{WeinbergCriterium} 
S. Weinberg and J. G. Taylor, 
Phil. Trans. R. Soc. London A {\bf 310}, 249--252 (1983).


\bibitem{Trialogue} 
M.J. Duff, L.B. Okun and  G. Veneziano,
JHEP 0203 (2002) 023.

 \bibitem{OkunCube} 
 L.B. Okun,    
 Cube or hypercube of natural units,   
 in: ``Multiple facets of quantization and supersymmetry'', Michael Marinov Memorial Volume, 
 Eds. M. Olshanetsky and A. Vainshtein, 
 World Scientific, 2002.

\bibitem{OkunGamov} 
L.B. Okun,
Physics of Atomic Nuclei {\bf 65}, 1370--1372 (2002). 

\bibitem{Gamov}
G. Gamow, D. Ivanenko and L. Landau,
 Physics of Atomic Nuclei  {\bf 65}, 1373--1375 (2002).

\bibitem{Klinkhamer2022b}
F.R. Klinkhamer,
Phys. Rev. D {\bf 106}, 124015 (2022).

\bibitem{Vergeles2022} 
S.N. Vergeles,
Class. Quantum Grav. {\bf 39}, 038001 (2022).

\bibitem{NiehYan1982a}
H.T. Nieh and M.L. Yan, 
 J. Math. Phys. {\bf 23}, 373  (1982).

 \bibitem{NiehYan1982b}
H.T. Nieh and M.L. Yan, 
Ann. Phys. {\bf 138}, 237 (1982).

\bibitem{Nieh2007}
H.T. Nieh,
Int. J. Mod. Phys. A {\bf 22},  5237 (2007).

\bibitem{Carlip2022} 
S. Carlip,
arXiv:2209.14282

\bibitem{Adler2022} 
S.L. Adler,
arXiv:2209.02537.

\bibitem{Adler2022b} 
S.L. Adler,
Int. J. Mod. Phys. D {\bf 31}, 2250070 (2022).

\bibitem{Klinkhamer2017}
F.R. Klinkhamer,
Int. J. Mod. Phys. D {\bf 26}, 1750006 (2017).

\bibitem{Klinkhamer2022}
F.R. Klinkhamer,
arXiv:2207.03453 [hep-th].
 

\end{thebibliography}
\end{document}